\documentclass[
prl,
reprint,
onecolumn,
nodate,
superscriptaddress,
 amsmath,
 amssymb,
 aps,
]{revtex4-2}

\pdfoutput=1

\usepackage[utf8]{inputenc} 
\usepackage{xcolor}
\usepackage{graphicx}
\usepackage{dcolumn}
\usepackage{bm}
\usepackage{microtype}
\usepackage{threeparttable} 
\usepackage{mathtools}
\usepackage[colorlinks=true]{hyperref}

\newcommand{\rfe}[1]{Eq.~(\ref{#1})}

\newcommand{\rff}[1]{Fig.~\ref{#1}}

\newcommand{\opd}{\hat{O}_{x}^{(5)}}
\newcommand{\opnd}{\hat{O}_{x}^{(6)}}

\newcommand{\be}{\begin{eqnarray}}
\newcommand{\ee}{\end{eqnarray}}

\newcommand{\pa}{{\cal A}}
\newcommand{\eL}{{\cal L}}
\newcommand{\eM}{{\cal M}}

\newcommand{\mysection}[1]{{\vspace{10 pt}\noindent \emph{{\textbf{#1}}--}}}

\begin{document}

\author{In\^{e}s Aniceto}
\email{I.Aniceto@soton.ac.uk}
\affiliation{School of Mathematical Sciences, University of Southampton, Southampton, UK}

\author{Jorge Noronha}
\email{jn0508@illinois.edu}
\affiliation{Illinois Center for Advanced Studies of the Universe \& Department of Physics, University of Illinois Urbana-Champaign, Urbana, IL 61801, USA}

\author{Micha\l\ Spali\'nski}
\email{michal.spalinski@ncbj.gov.pl}
\affiliation{National Centre for Nuclear Research, 02-093 Warsaw, Poland}
\affiliation{Physics Department, University of Bia{\l}ystok,
  15-245 Bia\l ystok, Poland}

\title{
    An analytic approach to the RTA Boltzmann attractor
}

\begin{abstract}
We reformulate the Boltzmann equation in the relaxation time approximation undergoing Bjorken flow in terms of a novel partial differential equation for the generating function of the moments of the distribution function. This is used to obtain an approximate analytic description of this system's far-from-equilibrium attractor via a series expansion at early times. This expansion possesses a finite radius of convergence and can be analytically continued to late times. We find that this procedure reproduces the known values of shear viscosity and other transport coefficients to high accuracy.  
We also provide a simple approximate analytic expression that describes the attractor in the entire domain of interest for studies of quark-gluon plasma dynamics. 
\end{abstract}

\maketitle

\mysection{Introduction} Studies of Bjorken flow \cite{Bjorken:1982qr} in models of quark-gluon plasma dynamics have led to
important insights into the physics of relativistic non-equilibrium evolution.
These include the significance of nonhydrodynamic modes and the role of
far-from-equilibrium attractors~\cite{Heller:2015dha} (see also the reviews  \cite{Florkowski:2017olj, Jankowski:2023fdz}). 
An important class of relativistic models describing the onset of hydrodynamic behavior is formulated in the language of 
kinetic theory. The simplest of these are based on the Boltzmann equation in the
relaxation time approximation (RTA), with many studies devoted to the dynamics of
Bjorken flow in this system. This is also our focus here. 

A very concise way of expressing the dynamics of kinetic theory is in the form of an
infinite hierarchy of ordinary differential equations for a set of moments of
the distribution function \cite{Grad-1949}. There are various ways of defining appropriate moments for relativistic gases \cite{Israel:1979wp,Denicol:2012cn,Denicol:2016bjh,Denicol_Rischke_book}.
Since in this Letter we will focus exclusively on Bjorken flow, it will be convenient to adopt the definitions of Blaizot and Yan~\cite{Blaizot:2017ucy}, which take advantage of the special features of boost-invariant and transversely homogeneous dynamics. 
An important element of our analysis is to reformulate the Blaizot-Yan hierarchy in terms of a new set of dimensionless 
moments and a dimensionless time variable $w\equiv\tau/\tau_R$, where $\tau$ is the proper time and $\tau_R$ is the relaxation time. 
It is then straightforward to make
contact with earlier studies of attractors in systems undergoing Bjorken flow~\cite{Heller:2016rtz,Florkowski:2017olj,Jankowski:2023fdz}. 

The physics
of this system has been explored by truncating the hierarchy at some level $L$
and studying the resulting finite set of coupled ODEs \cite{Blaizot:2017ucy,Blaizot:2019scw, Blaizot:2020gql}. One may use these truncated
systems to study the physics of this theory and find that some important features
are well-accounted for~\cite{Blaizot:2019scw}. At the same time, these truncations
miss features of the dynamics, such as the large order behavior of the gradient
expansion and the physics at early times. In particular, they 
reproduce the free streaming at early times only approximately and completely miss the properties of the large order behavior of the gradient expansion pointed out in Ref.~\cite{Heller:2016rtz}.

Our main goal in this paper is to pursue an alternative route by introducing a generating function for the moments. This function satisfies a
partial differential equation, which we derive in this Letter. 
This equation provides a new,  compact description of
the dynamics of this system; we study its solutions 
perturbatively in the early and late time regimes, which is equivalent to finding series expansions of the moments.

In the early time regime, we identify solutions that are regular at $w=0$ in the form of power series expansions with a non-zero radius of convergence. In particular, we calculate the series expansion of the pressure anisotropy of Bjorken flow --- this is the well-known far-from-equilibrium attractor of RTA kinetic theory. We show that the system is initially free-streaming, as expected on general grounds. We also find such solutions for the higher moments, which supports the claim that the entire distribution function follows a far-from-equilibrium attractor~\cite{Strickland:2018ayk}.   

We also study series solutions valid at late times, corresponding to the gradient expansion in hydrodynamics. The late-time solution of the PDE satisfied by the generating function leads to a very efficient method of generating the gradient expansion to very high order. The large-order behavior of this series extends the pattern of cuts in the Borel-Pad\'{e} plane noted in Refs.~\cite{Heller:2016rtz,Heller:2018qvh}. Since we generate many more terms of the gradient expansion, we also identify additional cuts. These features demonstrate the richness of RTA kinetic theory, showing that its early-time dynamics is very different from the Mueller-Israel-Stewart theory \cite{Muller:1967zza,Israel:1979wp} which  describes its near-equilibrium regime.

\mysection{The moment hierarchy} The RTA Boltzmann kinetic equation  for the case of Bjorken
flow takes the form 
\be
      \partial_\tau f(\tau,p)=\frac{f_{\rm eq}(\tau,p)-f(\tau,p)}{\tau_R}
\ee
where $f$ is the distribution function, $f_{\rm eq}$ is its equilibrium form (taken to be the Boltzmann distribution), $p$ is the $4$-momentum, and $\tau_R$ is the relaxation time.
We will consider 
an implicit time dependence for the relaxation time of the form $\tau_{R}= \gamma T\left(\tau\right)^{-\Delta}$,
where the effective temperature $T\left(\tau\right)$ is related to
the energy density via 
$\mathcal{E}\propto T^4$. 
Note that $\Delta=0$ corresponds to the case of constant relaxation
time \cite{Denicol:2019lio}, while $\Delta=1$ corresponds to the conformally invariant fluid~\cite{Heller:2016rtz,Heller:2018qvh}.

We follow Ref.~\cite{Blaizot:2017ucy}, which defines a set of moments of the distribution function
\be
\eL_n \equiv\int\frac{d^3p}{(2\pi)^3} p_0 \ P_{2n}(\cos \psi) \ f(\tau, p)~,
  \qquad \forall n \geq 0
\ee
where $\cos\psi=p_z/p_0=v_z$ is the particle velocity along the $z-$axis,
  and $P_{2n}$ are Legendre polynomials of degree $2n$.
Note that $\mathcal{E}\equiv\mathcal{L}_{0}$. The moments $\eL_n$ obey a hierarchy of coupled, ordinary differential equations
of the form
\begin{equation}
\tau\frac{d\mathcal{L}_{n}}{d\tau}=-a(n)\mathcal{L}_{n}-b(n)\mathcal{L}_{n-1}-c(n)\mathcal{L}_{n+1}-\frac{\tau}{\tau_{R}}\mathcal{L}_{n}\left(1-\delta_{n,0}\right)\;,\quad n\ge0\,.
\end{equation}
\noindent
with
\begin{align}
a(n) & =\frac{7}{4}+\frac{5}{16}\frac{1}{4n-1}-\frac{5}{16}\frac{1}{4n+3};\nonumber \\
b(n) & =\frac{1}{4}+\frac{n}{2}-\frac{5}{16}\frac{1}{4n-1}-\frac{9}{16}\frac{1}{4n+1};\label{eq:coeffs-a-b-c}\\
c(n) & =-\frac{n}{2}+\frac{9}{16}\frac{1}{4n+1}+\frac{5}{16}\frac{1}{4n+3}.\nonumber 
\end{align}

We now introduce the dimensionless moments (similar to \cite{Bazow:2015dha,Bazow:2016oky,Denicol:2016bjh,Mullins:2022fbx})
\begin{equation}
\mathcal{M}_{n} \equiv \frac{\mathcal{L}_{n}}{\mathcal{L}_{0}}\:,\quad n\ge0.
\end{equation}
\noindent
Note that $\eM_0 = 1$, while $\eM_1$ is related to the pressure anisotropy of Bjorken flow  $\pa = -3 \eM_1$ (see e.g. Refs.~\cite{Florkowski:2017olj,Jankowski:2023fdz}).  
This definition is motivated by earlier studies of hydrodynamization.

For $n\ge1$, one then finds the following hierarchy of differential equations:
\begin{equation}
\hat{O}_{w}\mathcal{M}_{n}+a(n)\mathcal{M}_{n}+b(n)\mathcal{M}_{n-1}+c(n)\mathcal{M}_{n+1}\:,\quad n\ge1,
    \label{eq:Master-moment-eq}
\end{equation}
\noindent
where $\hat{O}_{w}$
is independent of $n$ and given by 
\begin{align}
\hat{O}_{w} &
    =\left(1-\frac{\Delta}{3}-\frac{\Delta}{6}\mathcal{M}_{1}\right)w\frac{\partial}{\partial
    w}+\left(-\frac{4}{3}-\frac{2}{3}\mathcal{M}_{1}+w\right)\:.\label{eq:w-operator-Mn}
\end{align}
Note that to recover the energy density $\mathcal{E}\equiv\eL_0$, one 
also needs the relation 
\begin{equation}\label{eq:relation-M1-L0}
\left(1-\frac{\Delta}{3}-\frac{\Delta}{6}\mathcal{M}_{1}\right)\frac{\partial\ln\mathcal{L}_{0}}{\partial\ln w}+\frac{4}{3}+\frac{2}{3}\mathcal{M}_{1}=0.
\end{equation}
These equations constitute a reformulation of the original Boltzmann equation in the RTA.

\mysection{Generating functions for the moments} We now introduce a generating function for the moments,
\begin{equation}
    \label{eq:genfunc}
G_{\mathcal{M}}\left(x,w\right)=\sum_{n=0}^{+\infty}x^{n}\,\mathcal{M}_{n}\left(w\right),
\end{equation}
where $x$ is a formal variable.
This generating function satisfies a partial differential equation, which will be the basis of our subsequent investigations. To determine it we follow the general approach of Ref.~\cite{Krook-Wu} (see also \cite{Bazow:2015dha,Bazow:2016oky}): we 
multiply \rfe{eq:Master-moment-eq}
by $x^{n}$ and sum over all $n\ge0$, obtaining
\begin{align*}
\hat{O}_{w}G_{\mathcal{M}}\left(x,w\right)+\frac{7+x}{4}G_{\mathcal{M}}\left(x,w\right)-\frac{5}{16}\mathcal{M}_{0}+\frac{9}{16}\mathcal{M}_{1}+\frac{5}{16}\left(1+x\right)\sum_{n\ge1}\frac{x^{n}}{4n+3}\left(\mathcal{M}_{n+1}-\mathcal{M}_{n}\right)+\\
+\frac{9}{16}\sum_{n\ge1}\frac{x^{n}}{4n+1}\left(\mathcal{M}_{n+1}-\mathcal{M}_{n-1}\right)-\sum_{n\ge1}x^{n}\frac{n}{2}\left(\mathcal{M}_{n+1}-\mathcal{M}_{n-1}\right) & =0\,.
\end{align*}
We now change variables by setting $x=\xi^{4}$. A simple calculation reveals 
\begin{align*}
\frac{5}{16}\left(1+\xi^{4}\right)\sum_{n\ge1}\frac{\xi^{4n}}{4n+3}\left(\mathcal{M}_{n+1}-\mathcal{M}_{n}\right) & =\frac{5}{16}\frac{\left(1+\xi^{4}\right)}{\xi^{3}}\left[\int^{\xi}d\eta\,\left(\eta^{-2}-\eta^{2}\right)G_{\mathcal{M}}\left(\xi,w\right)-\mathcal{M}_{0}\int^{\xi}d\eta\,\eta^{-2}\right];\\
\frac{9}{16}\sum_{n\ge1}\frac{\xi^{4n}}{4n+1}\left(\mathcal{M}_{n+1}-\mathcal{M}_{n-1}\right) & =\frac{9}{16}\xi^{-1}\left[\int^{\xi}d\eta\,\left(\eta^{-4}-\eta^{4}\right)G_{\mathcal{M}}\left(\xi,w\right)-\int^{\xi}d\eta\,\left(\eta^{-4}\mathcal{M}_{0}+\mathcal{M}_{1}\right)\right];\\
-\sum_{n\ge1}\xi^{4n}\frac{n}{2}\left(\mathcal{M}_{n+1}-\mathcal{M}_{n-1}\right) & =-\frac{\xi}{8}\,\frac{d}{d\xi}\left[\left(\xi^{-4}-\xi^{4}\right)G_{\mathcal{M}}\left(\xi,w\right)-\xi^{-4}\mathcal{M}_{0}-\mathcal{M}_{1}\right]\,.
\end{align*}

To turn these differential/integral equations into a single
partial differential equation, we  need to multiply the full
equation by $\xi^{3}$ and then differentiate it $5$ times with respect to $\xi$, so that the indefinite integrals no longer appear.
Proceeding this way 
we find that 
\rfe{eq:Master-moment-eq} can be rewritten as 
the following PDE
\begin{equation}
    \left(
    \hat{O}_{w} \opd
    + \opnd
\right) 
G_{\mathcal{M}}\left(x,w\right)=0\,,
    \label{eq:Generating-function-equation}
\end{equation}
where
\be
\opd&=&\sum_{\ell=1}^{5}k_{\ell}\left(4\frac{\partial}{\partial\ln x}\right)^{\ell}\,,
\nonumber\\
\opnd&=&\sum_{\ell=0}^{6}\left(a_{-1}^{(\ell)}\frac{1}{x}+a_{0}^{(\ell)}+a_{1}^{(\ell)}x\right)\left(4\frac{\partial}{\partial\ln x}\right)^{\ell}.
\ee
\noindent
Here $k_{\ell}$ and $a_{i}^{(\ell)}$ are numerical coefficients
given by:
\begin{align}
    \{k_{\ell}\} &= \left\{ 0,-6,-5,5,5,1,0\right\} ;\\
\{a_{-1}^{(\ell)}\} &= \left\{ 0,60,-97,\frac{119}{2},-\frac{69}{4},\frac{19}{8},-\frac{1}{8}\right\} \,;\\
\{a_{0}^{(\ell)} \}&= \left\{ 0,-8,-5,10,\frac{35}{4},\frac{7}{4},0\right\} \,;\\
\{a_{1}^{(\ell)}\} &=\left\{ 1344,1928,1088,\frac{625}{2},\frac{97}{2},\frac{38}{8},\frac{1}{8}\right\} \,,
\end{align}
with $\ell\in\left\{ 0,1,\cdots,6\right\} $.
Note that the operator $\hat{O}_{w}$ 
appearing in \rfe{eq:Generating-function-equation} depends on $\eM_1 = \partial_x \left.G_{\mathcal{M}}\left(x,w\right)\right|_{x=0}$.

\mysection{Early time solutions}
\begin{figure}[t]
\centering
\includegraphics[width=6.5cm]{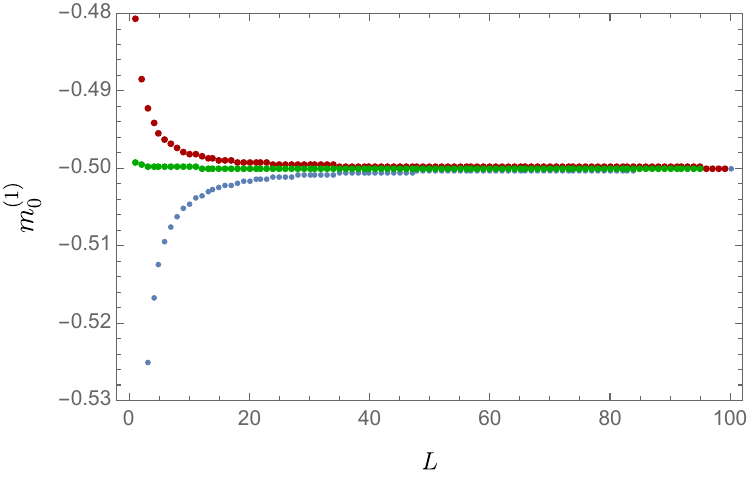}
\caption{Convergence-of the truncated calculation of coefficient $m^{(1)}_0$ as
    a function of truncation order $L$. Accelerated convergence via Richardson
    transforms is also shown (red - order 1; green - order 5).
    \label{fig:RT-for-early-time-coeff}}
\end{figure}
Our basic observation is that 
\rfe{eq:Generating-function-equation} can be used to calculate the
expansion of all the moments in powers of $w$. We will see that there are
exactly two such solutions regular at $w=0$, one of which is an attractor, while
the other is repulsive --- this is similar to what is known from studies of
MIS theory~\cite{Heller:2015dha,Aniceto:2022dnm}. The system also possesses solutions which are divergent at
$w=0$. Their series representation starts with $\eM_n = c_n/w^4 +\dots$, where the $c_n$'s are integration constants.  These are the generic solutions seen in numerical studies~\cite{Strickland:2018ayk}, which quickly approach the attractor already in the far-from-equilibrium regime, again in a way reminiscent of what is seen in MIS theory.  

We look for regular solutions by expanding the moments in powers of $w$:
\begin{equation}            
    \mathcal{M}_n=\sum_{k=0}^{+\infty}\,w^k\,m_k^{(n)}\quad,\qquad n\ge1.
\end{equation}
where $m_{0}^{(0)}=1$ and $m_{k}^{(0)}=0$ for $k\ge1$. 
The generating function can then be written as
\begin{equation}
G_{\mathcal{M}}\left(x,w\right)=\sum_{k\ge0}w^{k}\,h_{k}\left(x\right)\:,
    \quad
    \mathrm{where}\quad 
    h_{k}\left(x\right)\equiv\sum_{n=0}^{+\infty}x^{n}\,m_{k}^{(n)}\:.
    \label{eq:early-time-gen-func}
\end{equation}
The PDE given in \rfe{eq:Master-moment-eq} 
can be used to determine the expansion coefficients $m_{k}^{(n)}$ appearing above. Using it, one obtains the following relations:
\be
\frac{4}{3}\left(1+\frac{m_{0}^{(1)}}{2}\right)\opd h_{0}(x)-\opnd h_{0}(x)
&=&0\,;
\label{eq:heqnsa}\\ 
\left(\left(\frac{4}{3}+\Delta\frac{k}{3}\right)
\left(1+\frac{m_{0}^{(1)}}{2}\right) - k\right) \opd h_{k}(x) -
\opnd h_{k}(x) &=& \opd h_{k-1}(x) - \sum_{\ell=0}^{k-1}
\left(\frac{2}{3} + \Delta\frac{\ell}{6}\right) m_{k-\ell}^{(1)}\,\opd h_{\ell}(x).
\label{eq:heqnsb}
\ee
One first needs to find the initial condition $G_{\mathcal M}(x,0)=h_0(x)$,
which amounts to finding the initial values for the moments $m^{(n)}_0$ for
$n\ge 0$. This is determined by \rfe{eq:heqnsa},
which leads directly to the recursion relation 
\be
\label{eq:initialrr}
-K^{(n)}\,\left(\frac{4}{3} + \frac{3}{2} m_{0}^{(1)}\right) \,
m_{0}^{(n)}+A_{-1}^{(n+1)}\, m_{0}^{(n+1)}+A_{0}^{(n)}\, m_{0}^{(n)} +
A_{1}^{(n-1)}\,m_{0}^{(n-1)}=0\,,
\ee
where
\be
K^{(n)} &\equiv& \sum_{\ell=1}^{5}k_{\ell}\left(4n\right)^{\ell}\,;\\
A_{i}^{(n)} &\equiv& \sum_{\ell=0}^{6}a_{i}^{(\ell)}\left(4n\right)^{\ell}\;,\quad i=-1,0,1\,.
\ee
The recursion relations \rfe{eq:initialrr} possess two solutions; they can be obtained numerically by truncating the sequence at some level $L$ by setting $m_0^{(n)}=0$ for $n>L$. Here, we focus on the one that corresponds to the attractor. 
This way one finds that $m_0^{(1)} = -1/2$ (see \rff{fig:RT-for-early-time-coeff}).
Given this value, all the remaining  
$m_0^{(n)}$ can be generated from the recursion relation \rfe{eq:initialrr}. The resulting sequence turns out to be
\begin{equation}
m_{0}^{(n)}=\frac{(-1)^{n}}{2^{2n-1}}\frac{(2n-1)!}{n!\,(n-1)!}\,,\quad n\ge1,
\end{equation}
which translates into an exact form for the $w=0$ contribution to the generating function:
\begin{equation}
h_{0}(x)=\frac{1}{\sqrt{1+x}}.
\end{equation}
One can verify directly that this solves \rfe{eq:heqnsa} exactly. This is a very important result: it shows that there is a regular solution (attractor) for {\em all} the moments $\eM_n$. 
This demonstrates, for the first time, that the entire distribution function follows a far-from-equilibrium attractor already at early times, corroborating the observation made by Strickland in Ref.~\cite{Strickland:2018ayk} on the basis of numerical simulations.

\begin{figure}[t]
\centering
\includegraphics[width=6.5cm]{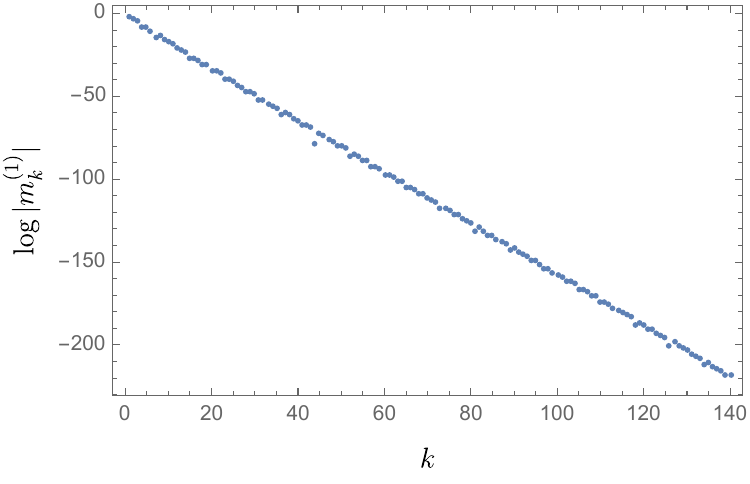}
\caption{Convergence of the early time expansion of the pressure anisotropy: this plot shows the exponential growth of the coefficients $m_k^{(1)}$ as a function of $k$. 
 \label{fig:Convergence-early-time}}
\end{figure}

The higher contributions $h_{k}(x), k > 0$ can be obtained by solving the
remaining 
recursion relations which follow from \rfe{eq:heqnsb}. 
Using the
fact that $m_{0}^{(1)}=-1/2$, we obtain 
\begin{align}
    K^{(n)}\,\left(k\left(1-\frac{\Delta}{4}\right)-1\right)\,m_{k}^{(n)}+A_{-1}^{(n+1)}
    \,m_{k}^{(n+1)} + A_{0}^{(n)}\,m_{k}^{(n)} + A_{1}^{(n-1)} \,m_{k}^{(n-1)}=\nonumber \\
    = -K^{(n)}\,m_{k-1}^{(n)}+K^{(n)} \,\sum_{\ell=0}^{k-1}\left(\frac{2}{3}+\Delta\frac{\ell}{6}\right)\,m_{k-\ell}^{(1)}\,m_{\ell}^{(n)}\,.
\end{align}
The form of the ODE solved by the function $h_{k}(x)$, \rfe{eq:heqnsb}, suggests a
decisive simplification. We 
set
\begin{equation}
h_{k}(x)=\sum_{\ell=0}^{k}\,\alpha_{k\ell}\,\tilde{h}_{\ell}(x)\,,\quad k\ge0
\end{equation}
where $\alpha_{kk}=1$, and $\tilde{h}_{0}=h_{0}$. 
By choosing the coefficients $\alpha$ such that the inhomogeneous
part of \rfe{eq:heqnsb} is canceled, we find that the functions $\tilde{h}_{k}$
obey the same ODE for each $k$:
\begin{equation}
\label{eq:odehtil}
\left(\frac{4-\Delta}{4}\,k-1\right)\opd\tilde{h}_{k}+\opnd \tilde{h}_{k}=0,\quad k\ge0.
\end{equation}
In order that this happens, the coefficients $\alpha_{k\ell}$ must obey 
\begin{equation}
\alpha_{k\ell}\,\frac{4-\Delta}{4}\left(k-\ell\right)=-\alpha_{k-1,\ell}+\frac{2}{3}\sum_{m=\ell}^{k-1}\,\frac{4+m\Delta}{4}\,m_{k-m}^{(1)}\,\alpha_{m\ell}\,,\label{eq:alpha-el-eqs}
\end{equation}
for $\ell=0,\cdots,k-1$, and $\alpha_{kk}=1$.
We can now write 
\begin{equation}
\tilde{h}_{k}=C_{k}\,\sum_{n=0}^{+\infty}\,x^{n}\,\beta_{k}^{(n)}\,,\quad \beta_{k}^{(0)}=1\,.
\end{equation}
in terms of new coefficients $\beta_{k}^{(n)}$, which can be determined recursively using \rfe{eq:odehtil}. 
This amounts to solving 
\begin{align}
    \left(K^{(n)}
    \,\left(k\,\frac{4-\Delta}{4}-1\right)+A_{0}^{(n)}\right)\,\beta_{k}^{(n)}+A_{-1}^{(n+1)}\,\beta_{k}^{(n+1)}+A_{1}^{(n-1)}\,\beta_{k}^{(n-1)}
    & =0\,,\quad n\ge1;\\
\beta_{k}^{(0)} & =1\,.
\end{align}
We proceed by truncating these relations at some level $L$ by setting 
$\beta_{k}^{(m)}=0$ for $m > L$.  This leads to a 
a system of linear equations of the form  
$Q\,\boldsymbol{x}+\boldsymbol{b}=0$ with a tridiagonal $L\times L$ matrix $Q$, a vector of unknowns $\boldsymbol{x}=\left(\beta_{k}^{(1)},\cdots,\beta_{k}^{(N)}\right)$
and a constant vector $\text{\ensuremath{\boldsymbol{b}}}=\left(A_{1}(0),0,\cdots,0\right)$, which can be solved very efficiently for any value of $k$. 
The results we show below were obtained for 
$L=220$. Increasing the number of moments included in the truncation will increase the accuracy of the coefficients $m_k^{(1)}$ of the pressure anisotropy (for any $k$). This calculation is very efficient: determining the coefficients for $k\le 160$ of the first $L=220$ moments takes less than 2 minutes in Mathematica. 
The large $k$ behavior of  the coefficient
$\beta_{k}^{(1)}$, which is related to the pressure anisotropy,
is given by 
\begin{equation}
\beta_{k}^{(1)}\sim\frac{32}{45}\frac{1}{k}+\mathcal{O}\left(k^{-2}\right),
\end{equation}
which does not change with truncation level for $L>200$. This is already a hint suggesting that the small $w$ expansion
of the pressure anisotropy possesses a finite radius of convergence. 

\begin{figure}
\centering
\includegraphics[width=6.5cm]{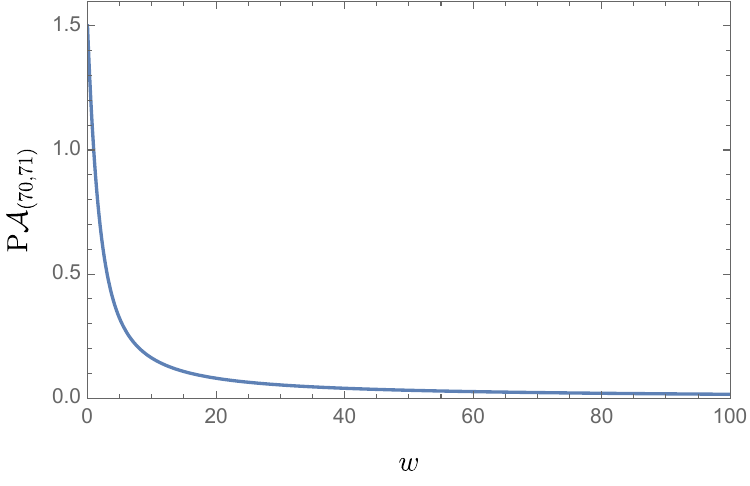}
\caption{Analytic continuation on early time expansion of the pressure anisotropy $\pa$ as a function of $w$, given by a Pad\'e approximant of order $(70,71)$. 
 \label{fig:Pade-earlyime}}
\end{figure}

The relation between the coefficients
$m_{k}^{(n)}$ of the moments and the $\beta_{k}^{(n)}$ is given 
by
\begin{equation}
\label{eq:mk1}
m_{k}^{(n)}=\sum_{\ell=0}^{k}\,C_{\ell}\,\beta_{\ell}^{(n)}\,\alpha_{k\ell}\,.
\end{equation}
Recalling that $\mathcal{M}_{0}=1$, and thus $m_{0}^{(0)}=1$, $m_{k\ge1}^{(0)}=0$,
we can determine the $C_{k}$ using  
\begin{equation}
\label{eq:Ck}
C_{k}=-\sum_{\ell=0}^{k-1}\,\alpha_{k\ell}\,C_{\ell}\,.
\end{equation}
The equations that determine the $\alpha_{k\ell}$ directly are
given in (\ref{eq:alpha-el-eqs}) for $\ell=1,\cdots,k-1$, recalling
that $\alpha_{kk}=1$. For $\ell=0$, we use (\ref{eq:alpha-el-eqs}) together with \rfe{eq:mk1}, which yields 
\begin{align}
\alpha_{k0}\left(\frac{4-\Delta}{4}k+\frac{2}{3}\,C_{0}\,\left(\beta_{k}^{(1)}-\beta_{0}^{(1)}\right)\right) & =-\alpha_{k-1,0}+\frac{2}{3}\sum_{r=1}^{k-1}\,\alpha_{kr}\,C_{r}\left(\beta_{r}^{(1)}-\beta_{k}^{(1)}\right)+\\
 & +\frac{2}{3}\sum_{m=1}^{k-1}\,\frac{4+m\Delta}{4}\,\alpha_{m0}\,\sum_{r=0}^{k-m}\,C_{r}\,\beta_{r}^{(1)}\,\alpha_{k-m,r}\,.\nonumber 
\end{align}
Note that the $\alpha_{k\ell}$ depend on no other coefficients of the functions $\tilde{h}_{k}$  except $\beta_{k}^{(1)}$. 
To solve them for each $k$, we need to start with
the $\alpha_{k\ell}$ with $\ell$ decreasing from $k$ to $0$. We can then find $C_k$ from \rfe{eq:Ck} and determine the moment coefficients in \rfe{eq:mk1}.

We now focus on the conformal case, $\Delta=1$. We have calculated the first $220$ moments, which
enables us to estimate the radius of convergence of the early-time expansion. In the case of the first moment (which is proportional to the
pressure anisotropy), one can easily see that the radius of convergence of the
small $w$ series of $\mathcal{M}_{1}$ follows from
$m_{k}^{(1)}\sim\mathrm{e}^{-a\,k}$ with $a\approx 1.55$. This can be seen in  \rff{fig:Convergence-early-time}.

We can now plot the pressure anisotropy $\pa=-3\mathcal{M}_{1}$ using the
convergent small $w$ expansion. To go beyond the radius of convergence (given by
$\mathrm{e}^{-a}$) we use analytic continuation by means of an off-diagonal
Pad\'e approximant (one order higher in the denominator than in the numerator, to account for expected behavior at large $w$). The result can be seen in 
\rff{fig:Pade-earlyime}, using a Padé approximant of order $(70,71)$, $\mathrm{P}\pa_{{70,71}}$. 
While this series solution was obtained at early times, the large radius of convergence makes it possible for this approximation of the far-from-equilibrium, pre-hydrodynamic attractor to be extended \emph{all the way} into the near-equilibrium domain. It is, therefore, very interesting to compare it to expectations based on hydrodynamics.
At $w\gg1$, hydrodynamics predicts the asymptotic behavior~\cite{Heller:2016rtz,Florkowski:2016zsi}
\be
\pa \sim \frac{8/5}{w}+\frac{32/105}{w^2}\,.
\ee
The leading term above reflects the known value of the shear viscosity for
RTA kinetic theory, $\eta/s = \gamma/5$ (see e.g.~\cite{Florkowski:2017olj}). Our early-time series solution analytically continued to $w = 100$
reproduces this value of the shear viscosity to three decimal places. 
This can be seen by 
re-expanding the Padé approximant
of the pressure anisotropy
$\mathrm{P}\pa_{(70,71)}$ for $w\gg 1$:
\be
\mathrm{P}\pa_{(70,71)}\sim \frac{1.60004}{w}+\frac{0.27348}{w^2},
\ee
which demonstrates the excellent agreement with the hydrodynamic behavior at large $w$. 

The validity of the analytic continuation over such a large domain 
is due to the large radius of convergence 
of the early-time expansion, along with the fact that singularities of the Pad\'e approximant appear in the second and third quadrants of the complex plane. We note that the same is not true for the generic solutions divergent at the origin.

We close this section with 
the following 
approximate analytic formula for the attractor in the form of an off-diagonal Pad\'e approximant 
\be
\pa = \frac{1500 - 12 w + 30 w^2}{1000 + 434 w + 95 w^2 + 9 w^3}
\ee
The values of the Pad\'{e} coefficients have been approximated to obtain the above expression. This formula works very well for the physically interesting range $0\le w \le 5$, i.e., from the earliest times all the way into the near-equilibrium regime. In particular, it reproduces the leading two orders of the early-time expansion, including the free-streaming behavior at $w=0$, and gives an accurate approximation up to $w\approx 5$ with a relative error of about $2\%$.

\begin{figure}[t]
\centering
\includegraphics[width=4.5cm]{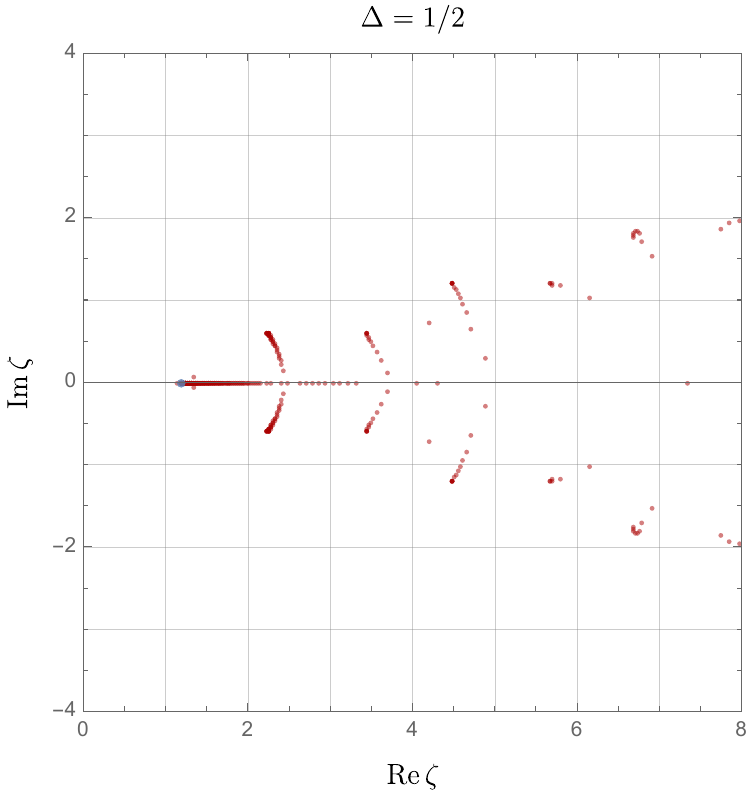}\hspace{15pt}
\includegraphics[width=4.5cm]{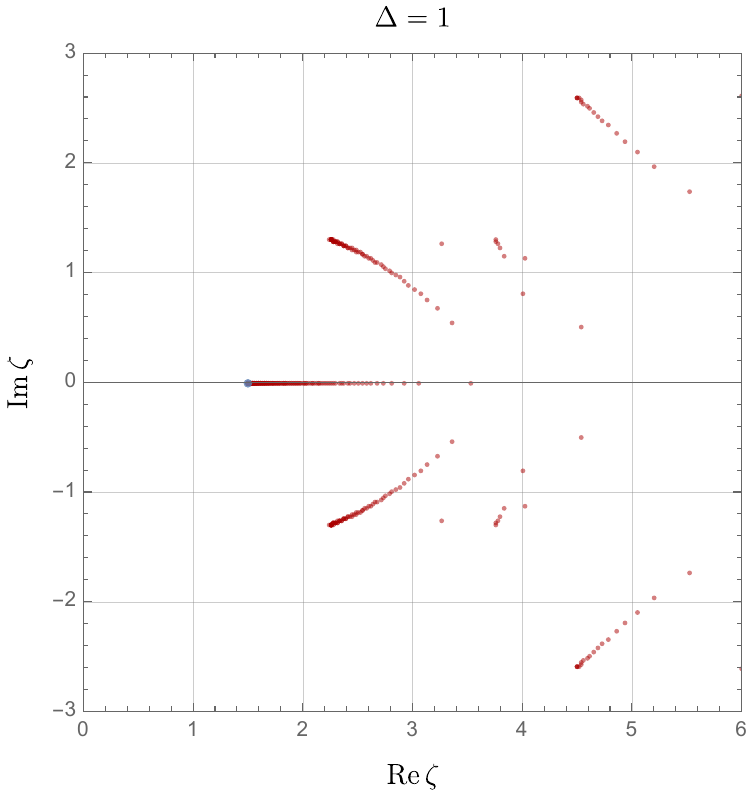}\hspace{15pt}
\includegraphics[width=4.5cm]{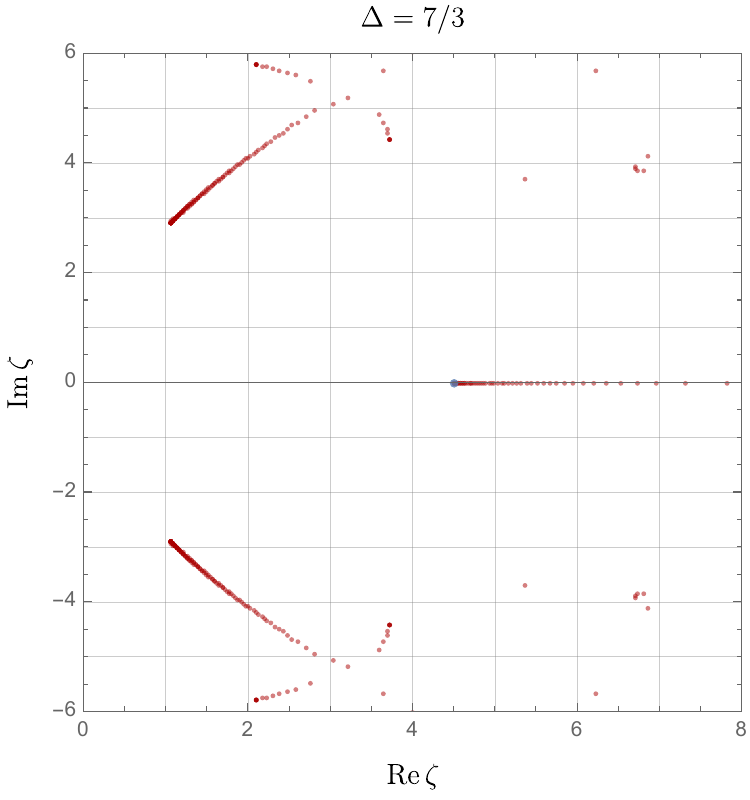}

\caption{Poles of the Borel-Pad\'{e} approximant of order 500 of the late time expansion of
the pressure anisotropy $\mathcal{A}$, $\mathrm{BP}_{500}(\mathcal{A})(\zeta)$, for different values of $\Delta$. In blue: start of the leading cut in the real axis, at position $\zeta_\Delta$: $\zeta_{1/2}=6/5$, $\zeta_{1}=3/2$ and $\zeta_{7/3}=9/2$.
 \label{fig:Borel-pade-poles-anisotropy}}
\end{figure}

\mysection{The late time expansion} The generating function approach that we have developed here can also be applied
to investigate the late-time expansion --- the asymptotic series for large
values of $w$:
\begin{equation}
\mathcal{M}_{n}=w^{-n}\sum_{k\ge0}w^{-k}M_{k}^{(n)}.\label{eq:Late-time-Mn}
\end{equation}
The PDE satisfied by the generating function, \rfe{eq:Generating-function-equation},  can be used to establish recursion relations for the coefficients $M_{k}^{(n)}$, in a very similar way to what was presented in the previous Section. The main difference is that changing the truncation at a particular order $w^{-k}$ does not change the accuracy of the lower orders 
already calculated. The recursion relations can be used to generate these coefficients in a vastly more efficient way than by using hitherto existing methods: we have been able to calculate $1000$ of them in about $10$ minutes. In contrast to the early-time solutions, the series appearing here have zero radius of convergence and need to be interpreted in the sense of  asymptotic analysis. 
The series coefficients grow factorially, so it is natural to use a Borel transform to give meaning to the divergent sum. The analytic continuation of the Borel transform, which can be carried out by means of a Pad\'{e} approximant,
reveals a pattern of singularities whose physical significance is of great interest. Here we present some results concerning the pressure anisotropy $\pa$, related to the moment $\mathcal{M}_1$, shown in Fig.\
\ref{fig:Borel-pade-poles-anisotropy}. 
The singularity structure is
strongly dependent on the value of the parameter $\Delta$, as noted in Ref.~\cite{Heller:2018qvh}. 

In the conformal case ($\Delta=1$), we reproduce the pattern found in Ref.~\cite{Heller:2016rtz}, consisting of the cut on the real axis and a symmetric pair of branch points with nonvanishing imaginary parts. 
Since we can easily generate many more terms of the gradient expansion than were hitherto available, we can also discern further branch points at complex-conjugate locations off the real axis. It is natural to conjecture that there is an infinite number of such cuts. While the cut on the real axis clearly corresponds to the nonhydrodynamic mode of the RTA kinetic theory, to identify the physical meaning of the singularities with nonvanishing imaginary parts remains a challenge for the future.

\mysection{Outlook} In this Letter, we studied the dynamics of the RTA Boltzmann equation by applying generating function techniques to the hierarchy of moment equations.
This yields a new
reformulation of this theory in terms of a partial differential equation in 
two
variables: the formal variable $x$, whose power tags each of the moments, and $w$, whose power at large times encodes the order of the gradient expansion.
We expect that several further insights can be obtained from this approach. 

Here, we have focused mainly on the early-time behavior of RTA kinetic theory.
Earlier works showed that the dynamics of this model reaches the
hydrodynamic domain following a far-from-equilibrium attractor, which has previously been obtained by numerical solutions of the Boltzmann equation. 
Using the generating function technique presented here, we have provided the first solid analytic evidence for
this attractor. We have, in particular, obtained a series solution with a finite
radius of convergence, which provides an accurate account of the attractor in a
large domain, fully describing the passage from the far-from-equilibrium regime
to the near-equilibrium domain. 

We have also shown that the formulation of RTA kinetic theory presented here opens the door to calculations of the gradient expansion to large orders at a qualitatively new level of 
efficiency, allowing for hundreds of coefficients to be computed in minutes (as opposed to weeks). This may eventually lead to a better understanding of the peculiar features
hinted at by the results of earlier studies of this problem. 
It would be interesting to see how this generating function technique 
can lead to new insight into attractors in other kinetic systems, such as $\lambda\phi^4$ theory \cite{Denicol:2022bsq} and high-temperature QCD plasmas \cite{Almaalol:2020rnu}.

\begin{acknowledgments}
\mysection{Acknowledgements} IA and JN thank KITP Santa Barbara for its hospitality during “The
Many Faces of Relativistic Fluid Dynamics” Program, where part of this work was conducted. 
IA and MS thank the Isaac Newton Institute for hosting them during the program "Applicable Resurgent Asymptotics" during the early stages of the work. IA has been supported by the UK EPSRC Early Career Fellowship EP/S004076/1. MS was supported by the National Science Centre, Poland, under Grant No. 2021/41/B/ST2/02909.  JN is partly supported
by the U.S. Department of Energy, Office of Science, Office for Nuclear Physics under Award No. DE-SC0023861. This research was partly supported by the National Science Foundation under Grant No. NSF PHY-1748958. 

\end{acknowledgments}

\bibliographystyle{JHEP}

\bibliography{rta.bib,References_Jorge}

\end{document}